\input aa.cmm 
\input psfig.tex
\voffset=1.0truecm
\def\gr{$\gamma$-ray }
\def\grs{$\gamma$-rays }
\MAINTITLE={A spectral study of gamma-ray emitting AGN}
\AUTHOR={M. Pohl@{1}, R.C. Hartman{@2}, B.B. Jones@{3}, P. Sreekumar@{2,4}}
\INSTITUTE={@1 Max-Planck-Institut f\"ur Extraterrestrische Physik,
Postfach 1603,
85740 Garching, Germany
@2 NASA/Goddard Space Flight Center, Code 660, Greenbelt, MD 20771, USA
@3 W.W. Hansen Expt. Physics Lab., Stanford Univ., Stanford, CA 94305-4085, USA
@4 Universities Space Research Association, USA}
\OFFPRINTS={\hphantom{http://www.gamma.mpe-garching} http://www.gamma.mpe-garching.mpg.de/$\sim$mkp/mkp.html}
\DATE={ }
\ABSTRACT={In this paper we present a statistical analysis of the
$\gamma$-ray spectra of flat-spectrum radio quasars (FSRQ) compared to those
of BL Lacs. The average spectra and possible systematic deviations from 
power-law behaviour are investigated by summing up
the intensity and the power-law fit statistic for both classes of objects.
We also compare the time-averaged spectrum to that
at the time of \gr outbursts.

The spectrum of the average AGN is softer than that
of the 
extragalactic $\gamma$-ray background. It may be that BL Lacs, which
on average have a harder spectrum than FSRQs, make up the bulk of the 
extragalactic background.

We also find apparent cut-offs at both low and high energies in the spectra
of FSRQs at the time of $\gamma$-ray
outbursts. While the cut-off at high energies may have something to do
with opacity, the cut-off at low energies may be taken as indication that
the $\gamma$-ray emission of FSRQs is not a one component spectrum.
}

\KEYWORDS={Methods: statistical; BL Lacertae objects: general; quasars: general; Gamma rays: observations}

\THESAURUS={03.13.6; 11.02.1; 11.17.3; 13.07.2}

\maketitle

\titlea{Introduction}

Up to the end of 1994 more than 50 extragalactic radio sources were
detected as emitters of high-energy \grs by EGRET. The majority of the
sources are flat-spectrum radio quasars (FSRQ) and about 20\% 
are classified as BL Lacertae (BL Lac) objects. Especially 
in their flare state FSRQs often
emit the bulk of their luminosity in the form of \grs (von Montigny et al.
1995). For a number of objects COMPTEL data at a few MeV (Sch\"onfelder
et al., 1996), OSSE data at a few hundred keV
(McNaron-Brown et al. 1995) or the extrapolation of X-ray data imply
a spectral break at a few MeV photon energy. This break may be attributed to
incomplete cooling of radiating electrons (Dermer and Schlickeiser 1993,
Sikora et al. 1994) or bremsstrahlung and annihilation emission of cooling
pair plasma (B\"ottcher and Schlickeiser 1995).

Individual \gr spectra in the EGRET range, i.e. above 30 MeV, can generally be
well described by power-laws. For the BL Lac Mrk421 the power-law
behaviour extends up to TeV energies (Punch et al. 1992).
Though a number of papers deal with the multifrequency spectrum of
\gr AGN, to our knowledge no attempt has yet been made to investigate whether
there are systematic deviations from power-law behaviour in the \gr spectra
of AGN. Due to the different environmental conditions one might expect
a different impact of opacity effects, for example, on FSRQs on
one side and on BL Lacs on the other side.

In this paper we analyse the class-averaged spectra of AGN by summing
the observed intensity and the statistic of power-law fits to the observed
emission.
We also derive the spectrum of the average \gr AGN which is to
be compared to the spectrum of the diffuse extragalactic background.
In the next section (Sec.2) we consider the time-averaged behaviour of
sources by using the summed EGRET data of Phases 1,2 and 3 corresponding
to observation times between May 1991 and October 1994. Since many of the
extragalactic sources are variable at \grs we discuss the behaviour
at flare state in Sec.3. A discussion of the 
results is in Sec.4.

\titlea{The average source}
Here we have summed all EGRET data taken between May 1991 and October 1994
to obtain \gr allsky maps
of counts and exposure. A quality cut has been imposed by restricting the
field-of-view in the individual viewing periods to 30$^\circ$ off-axis
angles before summing. At energies above 100 MeV we have performed a
standard search for point sources using a maximum-likelihood technique
(Mattox et al., 1996). The data in the energy range between 30 MeV and
100 MeV are usually neglected in this task since they add little to the 
statistics and source positioning, but require the energy-dependent
point-spread function be averaged over an even larger energy range
with corresponding systematic uncertainties.
In total we found 148 excesses with a likelihood test
statistic $TS=(2\ln \lambda_0 -2\ln \lambda)$
of least 9, of which 44 are positionally coincident with AGN known to emit
\grs in the EGRET range.
Table 1 lists
these sources which can be further divided into BL Lacs on one side
(11 objects) and FSRQs on the other side (33 objects). 
All sources except one (2155-304) have been observed with likelihood
test statistic higher than 25 in individual viewing periods or specific
combinations thereof
(Thompson et al. 1995, 1996). The source 2155-304 is observed with
likelihood test statistic of 33 in a later viewing period (VP 404)
during Phase 4 of the CGRO observing program (Vestrand et al. 1996). 
We do not intend to find new AGN in our sample, but select the already
identified ones for further analysis. Our threshold criterion of $TS\ge 9$
thus corresponds to a detection significance of at least 3$\,\sigma$
in the time-averaged data.
Some  
AGN which have been found with high significance in individual viewing
periods fail to show up in Table 1 since they have been absent in other
viewing periods so that their average signal is below threshold. 

\begtabfull
\tabcap{1}{A list of radio sources which can be identified with point
sources in the summed EGRET data of the time period 5/91 to 10/94 at
energies above
100 MeV. The sources are categorized to belong either to the BL Lac class
or to the class of FSRQ. The second column indicates the EGRET reference
with acronyms for
the second EGRET catalogue (2EG, Thompson et al. 1995), its supplement (2EGS, 
Thompson et al. 1996), and the papers
by Mukherjee et al. (1997) (M) and Vestrand et al. (1996) (V).
If there is at least a moderate level of variability, the third column
gives the viewing period number for the peak flux in standard EGRET notation.
No entry in this column indicates no evidence for variability.
Note the following constraints for the selection of the viewing period of peak flux: the total statistical significance is required to be $\ge\,4\,\sigma$ and preference is given to data taken with
smaller aspect angle in case of viewing periods of comparable duration and
comparable source flux. }
$${\offinterlineskip \tabskip=0pt
\vbox{ \halign{\strut\hfil\vrule#&\ \hfil#\hfil\ &\vrule#&\ \hfil\ #\ \hfil&
\vrule#&\ \hfil\ #\ &\vrule#&\ \vrule#&\ \hfil\ #\ \ \hfil&
\vrule#&\ \hfil\ #\ \hfil&\vrule#&\ \hfil\ #\ &\vrule#\strut\cr
\noalign{\hrule}&FSRQ&&EGRET&&Peak&&&BL Lacs &&EGRET&&Peak&\cr
& && &&VP &&& && &&VP &\cr\noalign{\hrule}
&0202+149&&2EG&&21&&&0235+164&&2EG&&21&\cr
&0208-512&&2EG&&10&&&0521-365&&2EG&& &\cr
&0234+285&&2EG&&21&&&0537-441&&2EG&&6&\cr
&0420-014&&2EG&&21&&&0716+714&&2EG&& &\cr
&0440-003&&2EGS&&337&&&0735+178&&2EG&&40&\cr
&0446+112&&2EG&&36.5&&&0829+046&&2EG&& &\cr
&0528+134&&2EG&&213&&&0954+658&&2EG&&228&\cr
&0827+243&&2EG&& &&&1101+384&&2EG&& &\cr
&0836+710&&2EG&&22&&&1219+285&&2EGS&&308.6&\cr
&0917+449&&2EG&&326&&&1604+159&&2EG&&25&\cr
&0954+556&&2EG&&319&&&2155-304&&M,V&& &\cr
&1127-145&&2EG&&206&&& && && &\cr
&1156+295&&2EG&&206&&& && && &\cr
&1222+216&&2EG&&204&&& && && &\cr
&1226+023&&2EG&&308.6&&& && && &\cr
&1229-021&&2EG&& &&& && && &\cr
&1253-055&&2EG&&3&&& && && &\cr
&1406-076&&2EG&&205&&& && && &\cr
&1424-418&&2EGS&&314&&& && && &\cr
&1510-089&&2EG&&24&&& && && &\cr
&1606+106&&2EG&&16&&& && && &\cr
&1611+343&&2EG&&202&&& && && &\cr
&1622-253&&2EG&& &&& && && &\cr
&1633+382&&2EG&&9.2&&& && && &\cr
&1730-130&&2EG&&334&&& && && &\cr
&1739+522&&2EG&&212&&& && && &\cr
&1908-201&&2EG&&323&&& && && &\cr
&1933-400&&2EG&&35&&& && && &\cr
&2022-077&&2EG&&7.2&&& && && &\cr
&2052-474&&2EG&&42&&& && && &\cr
&2230+114&&2EG&& &&& && && &\cr
&2251+158&&2EG&&37&&& && && &\cr
&2356+196&&2EG&& &&& && && &\cr
\noalign{\hrule}}}}$$
\endtab

\vskip0.4truecm
{\psfig{figure=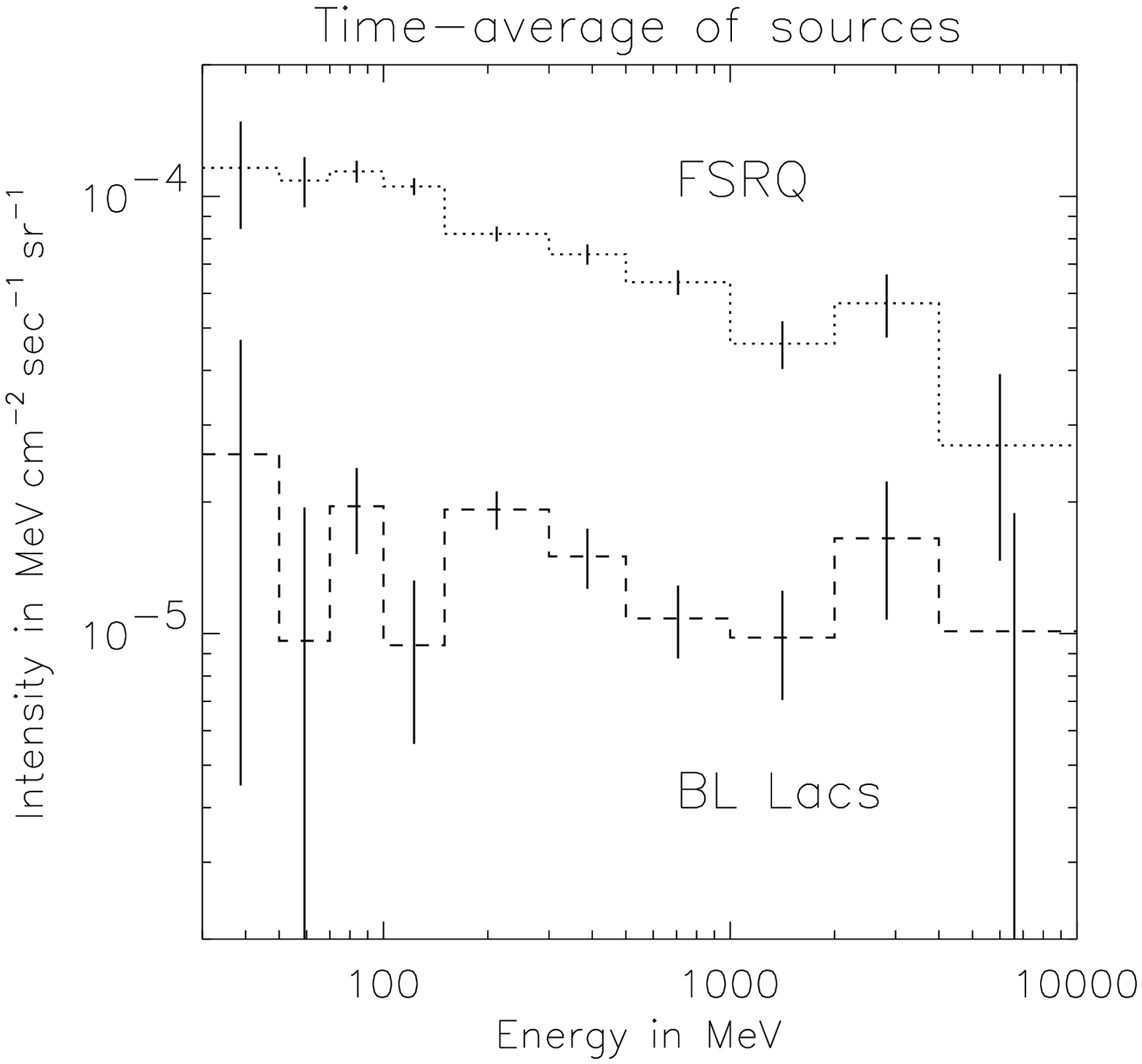,width=8.5cm,clip=}
\figure{1}{The summed intensity spectrum of the 33 identified FSRQ
and 11 identified BL Lacs
in the total EGRET allsky data above 100 MeV. The error
bars are derived by Gaussian error propagation of the individual uncertainty measures. The spectrum of FSRQ is significantly softer than that of the
diffuse background, while that of BL Lacs is consistent with it.}
\vskip0.2truecm}

For all 44 AGN we have performed a spectral analysis. The positions
of the \gr sources have been set to those of the corresponding radio
sources before doing a likelihood analysis of all 148 excesses in ten
energy bands. Since the best model for the data consists of diffuse emission
and 148 sources, we have to fit all these quasisimultaneously although
we are interested only in the results for the 44 AGN.
Observations of \gr pulsars have shown that the calibrated effective area
in the two low energy bins (30-50 MeV and 50-70 MeV)
is overestimated. In the standard analysis the observed source flux
is multiplied by a correction factor which accounts for the 
miscalibration. The accuracy of this correction factor can be estimated to
be better than 30\% at 30 MeV to 50 MeV and better than 10\% at 50 MeV to
70 MeV (Fierro 1995, $\S$3.4).

We have summed the observed intensity in the ten energy bands
to derive the spectrum of the average AGN. This spectrum is what we would
get as contribution to the diffuse 
extragalactic background if the AGN were unresolved. This `average' AGN
spectrum can not be directly compared to other studies in which
power-law fits of the spectra of individual AGN are averaged (e.g. Mukherjee
et al. 1997). However for the purpose of comparison with the diffuse
extragalactic background co-adding of the observed intensity spectra without
any normalization or fitting procedure is the appropriate method.
Since the radio catalogues
are incomplete near the galactic plane we do not expect -- and in fact we do
not get -- identifications for sources located at $|b| \le 10^\circ$.
Thus effectively there are only 
$\sim$10.4 steradian sky area which we have searched
for point sources and which is to be used to derive the corresponding
average \gr intensity. The results is shown in Fig.1 separately for FSRQs and
BL Lacs. 

The sum of the \gr intensity contribution of all 44 AGN is about 7\% of the
total diffuse background intensity.
It is interesting to see that the \gr spectrum
of the average BL Lac seems to be harder than that of the average FSRQ.
The difference in spectral index is $\delta s = 0.12\pm 0.08$ and is thus
not very significant which is mainly due to the large uncertainties in the
average BL Lac spectrum.
This does not imply that in single viewing periods BL Lacs have always
harder spectra than FSRQ. In fact we see a remarkable spread of
spectral indices for both classes of objects when individual
viewing periods are considered (Mukherjee et al., 1997).
We also know that individual sources can change
their spectrum: there is a trend that the spectrum hardens with increasing
flux level (M\"ucke et al. 1996). But this concerns individual sources.
The spectrum of the average in both object classes
appears to be different, and therefore they would
contribute with different spectral characteristic to the diffuse
\gr background. 

The spectrum of the average AGN is dominated by that of the FSRQs and it
differs significantly from that of the observed
diffuse extragalactic background (Kniffen et al. 1996; Sreekumar et al.,
1997) as we show in Fig.2.
When fitting the ratio of both spectra with a
power-law we find the background spectrum harder than that of
the average AGN by $\delta s =0.15\pm 0.04$ when all energy bins are
considered and by $\delta s =0.17\pm 0.045$ when only the data with
very good statistics between 70 MeV and 4 GeV are considered. The
goodness-of-fit for a constant is $8\cdot 10^{-3}$ and $3.5\cdot 10^{-3}$,
respectively, which corresponds to about 2.7$\sigma$ significance. A
Fischer-Snedecor-test indicates that with 3.5$\sigma$ significance
a linear relation is a better fit to the intensity ratio than a constant. 
Especially 
between 70 MeV and 4 GeV the uncertainty of the background intensity is
so small that the assumption of Gaussian statistics for the uncertainty of the
intensity ratio can be regarded as a reasonable approximation.
Since we see some objects
only at flare state when they tend to have harder spectra, the true average
spectrum of AGN may be even softer and thus further away from agreement
with the observed spectrum of the \gr background.

There is no cut-off visible in the \gr spectra with the possible
exception of a weak deficiency below 100 MeV for the FSRQs which
may be the outer extension of the usual roll-over at a few MeV
(Sch\"onfelder et al. 1996). However,
the expected spectral form of the average intensity is not necessarily
power-law.
Here we have summed over many sources with different spectra. Also some of
the individual sources have changed their \gr spectrum over the last
years. A summation over different power-law like spectra results in a
positive curvature of the average spectrum (Brecher and Burbidge 1972). 
Given the spread in
spectral indices for the 44 sources we can estimate that the averaging 
should change the spectral index by around 0.1 between 100 MeV
and 10 GeV. 

We have also searched for systematic deviations from power-law
behaviour in the \gr spectra of the 44 AGN. For each individual AGN
we have fitted a power-law spectrum to the data. The difference
between this fit and the measured intensity in the ten energy bands
weighted by the observational uncertainty,
i.e. $\chi = (I_{fit} -I)/(\delta I)$, has been summed for all $N$ FSRQ
and BL Lacs, respectively, to obtain the net deviation
$$\chi_{tot} = {{1}\over {\sqrt{N}}}\ \sum_{i=1}^N \chi_i\ \eqno(1)$$
where the applicability of Gaussian statistics has been assumed for
the renormalization ${{1}\over {\sqrt{N}}}$. 
 
\vskip0.4truecm
{\psfig{figure=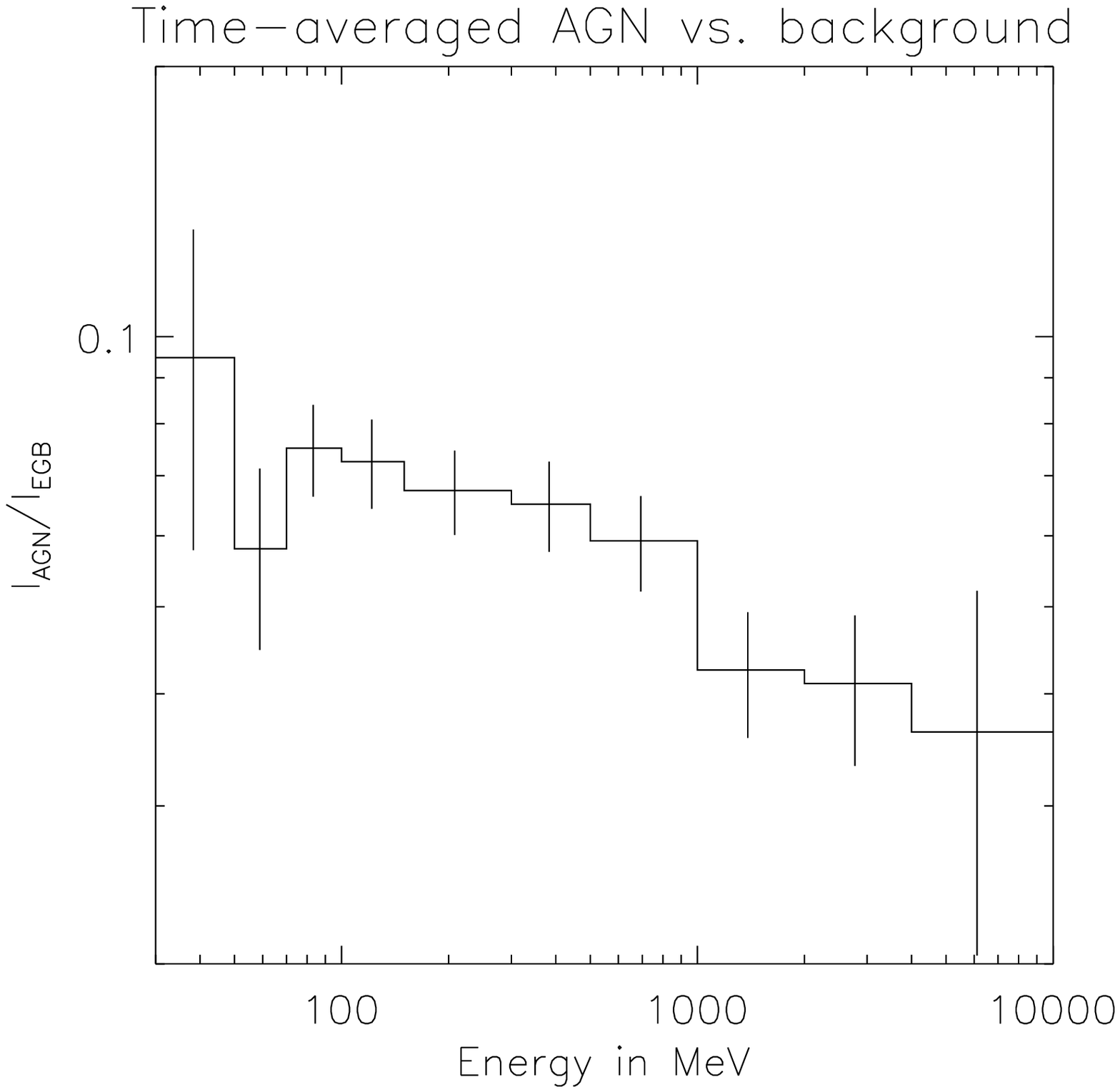,width=8.5cm,clip=}
\figure{2}{The ratio of the average intensity of all observed AGN to that of
the extragalactic diffuse \gr background. This ratio is not compatible with
a constant and thus the observed AGN would, if they were unresolved, give a
softer diffuse emission than we observe in the background. This implies that
the background can not be the superposition of unresolved AGN with the
same characteristics as the observed objects.}
\vskip0.2truecm}

We have verified with
Monte-Carlo simulations that there are no significant deviations from
a Gaussian distribution of the number $\chi_{tot}$.
The results are shown in Fig.3. No significant deviations
from power-law behaviour in the AGN spectra are observed,
either for FSRQ or for BL Lacs. The BL Lac fit statistic is noisy.

\vskip0.4truecm
{\psfig{figure=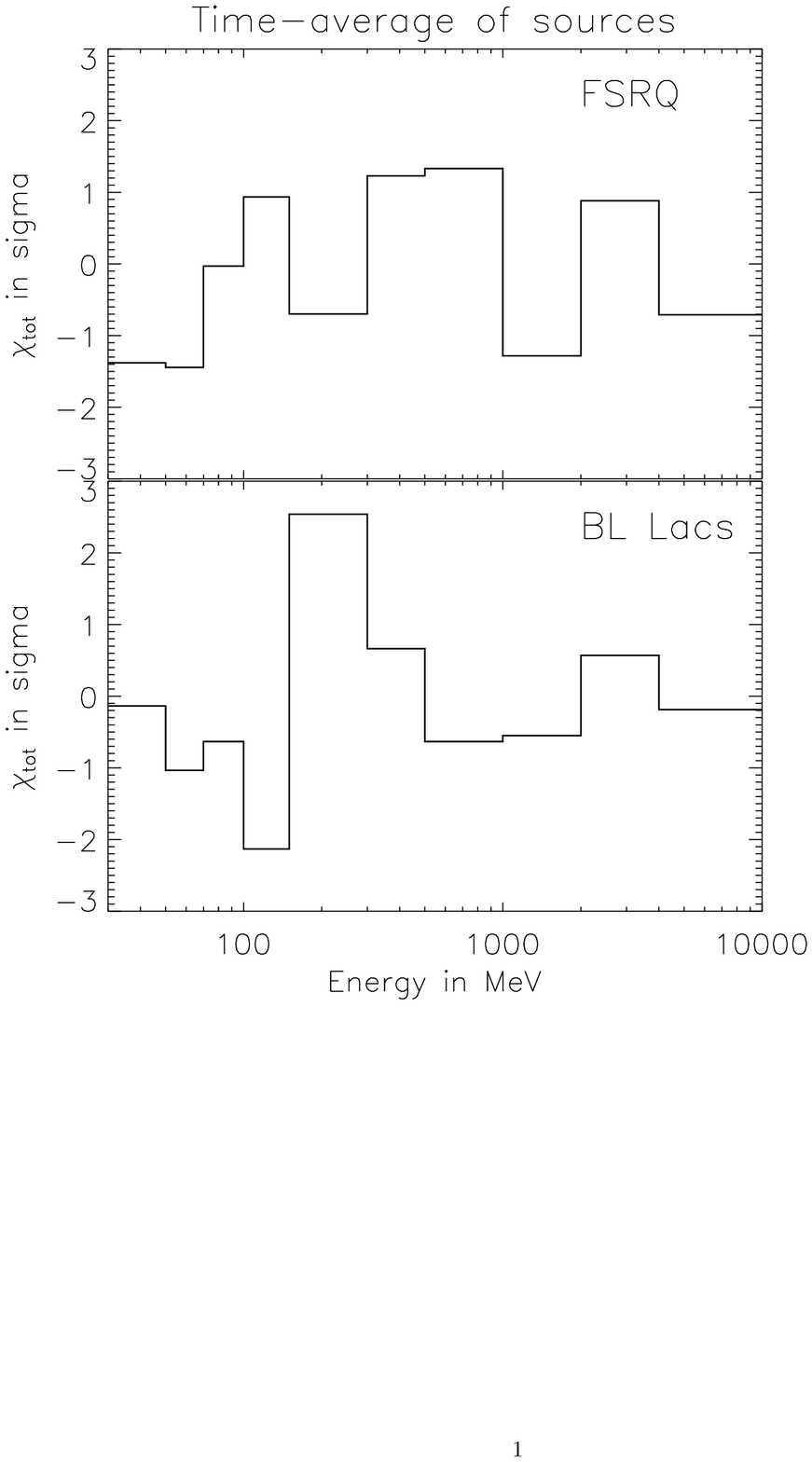,width=8.5cm,clip=}
\figure{3}{The summed power-law fit statistic of the 33 identified
FSRQ and 11 BL Lacs in the total EGRET allsky data
according to Eq.1. There is no significant deviation from
power-law behaviour in the average FSRQ and BL Lac spectrum.
The two extreme values with opposite sign between 100 MeV and 300 MeV
in the BL Lac result are unlikely to be real.}
\vskip0.2truecm}

\titlea{The peak spectra}

A large fraction of AGN is variable at \gr energies. Any cut-off arising from
opacity effects will be more prominent at high flux levels since then
the intrinsic photon density of the source is high. We have therefore chosen 
a subsample of AGN for which at least a moderate level of variability
can be found. These sources are indicated in Table 1 by the remark `yes'
in the variability column. The light curves of the individual sources have been
taken from the second EGRET catalogue (Thompson et al., 1995)
and from the standard viewing period analysis of the EGRET group at
MPE.
In total we are left with 28 FSRQs and only 6 BL Lacs which are variable.
The analysis is now similar to that described
in the previous section except that the spectra are not derived on the basis
of the summed data of Phases I-III but only on data of the viewing periods in
which the sources showed the highest flux levels. We should note that detections
with formal significance $\le\,4\,\sigma$ have not been considered. 
In case of two or more viewing periods of comparable duration and comparable
source flux preference has been given to data taken with smaller aspect angle.
The results for the 
summed spectra are presented in Fig.4 while the summed power-law
fit statistic is shown in Fig.5.

\vskip0.4truecm
{\psfig{figure=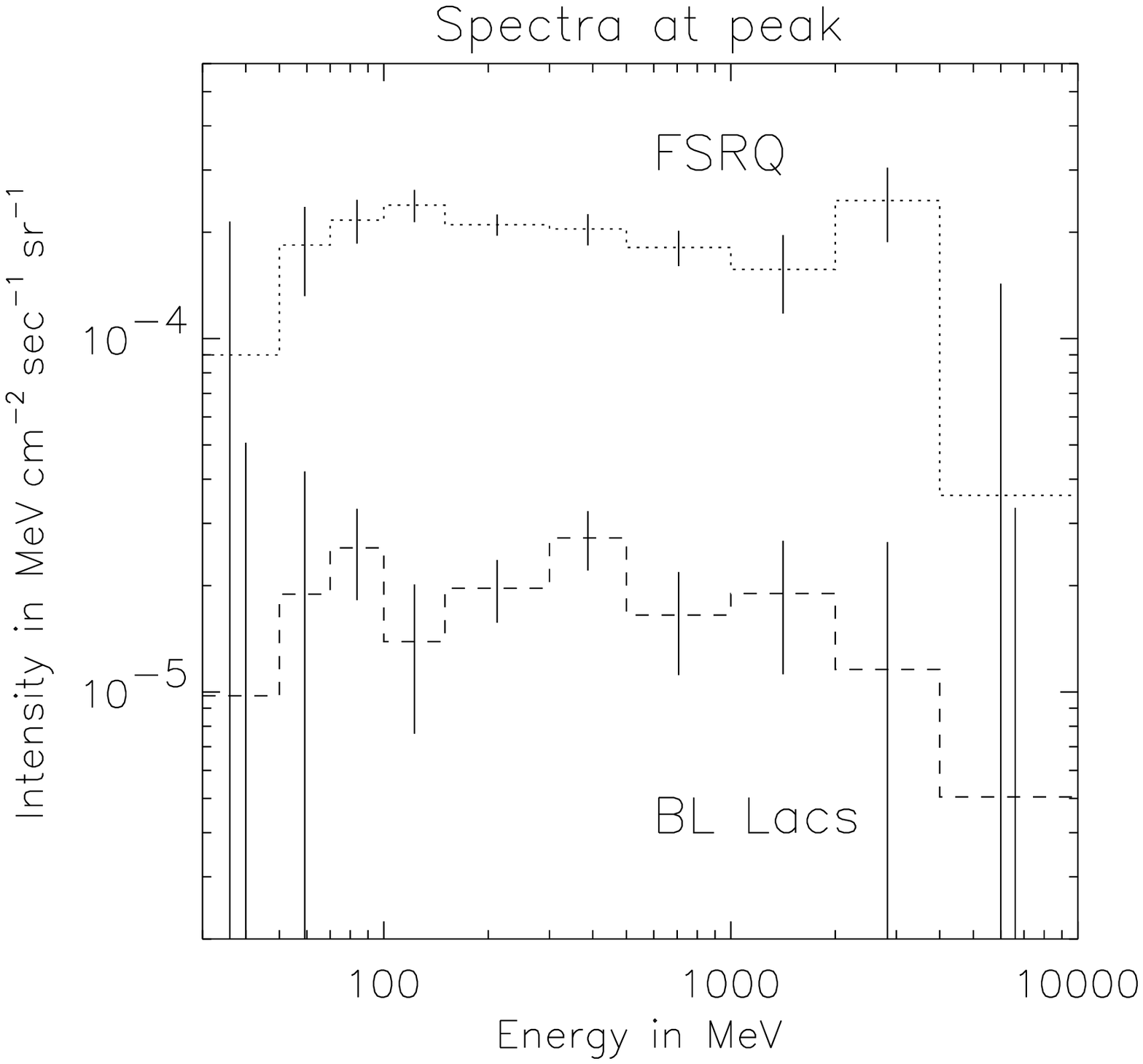,width=8.5cm,clip=}
\figure{4}{The summed intensity spectrum of the 28 variable FSRQ and
6 variable BL Lacs
at the time of the highest flux level above 100 MeV. The error
bars are derived by gaussian error propagation of the individual uncertainty measures. The FSRQ spectrum at peak is significantly harder than the
spectrum of the average FSRQ in Fig.1.}
\vskip0.2truecm}

There is not very much change compared to the average behaviour in
case of BL Lacs. One has to keep in mind that we are now left
with 6 objects and the statistics are not sufficient to distinguish
general trends from pathological individuals. 

The behaviour of FSRQs during their peak phase is more interesting.
At first we see that the flare spectra are harder than the average, at least
between 100 MeV and
a few GeV. This is a confirmation of a claim by M\"ucke et al. (1996)
who found a hardening of the \gr spectra with increasing flux level for 8 
highly variable AGN. When fitting a power-law to the intensity ratio of
average FSRQ at peak and total average FSRQ we find the spectrum
of the average FSRQ at peak harder than that of
the time-average by $\delta s =0.19\pm 0.05$ when all energy bins are
considered and by $\delta s =0.18\pm 0.06$ when only the data with
very good statistics between 70 MeV and 4 GeV are considered. The
goodness-of-fit for a constant is $2.5\cdot 10^{-3}$ and $7.5\cdot 10^{-3}$,
respectively, which corresponds to about 2.7$\sigma$ significance. A
Fischer-Snedecor-test indicates that with 4.0$\sigma$ significance
a linear relation is a better fit to the intensity ratio than a constant.

\vskip0.4truecm
{\psfig{figure=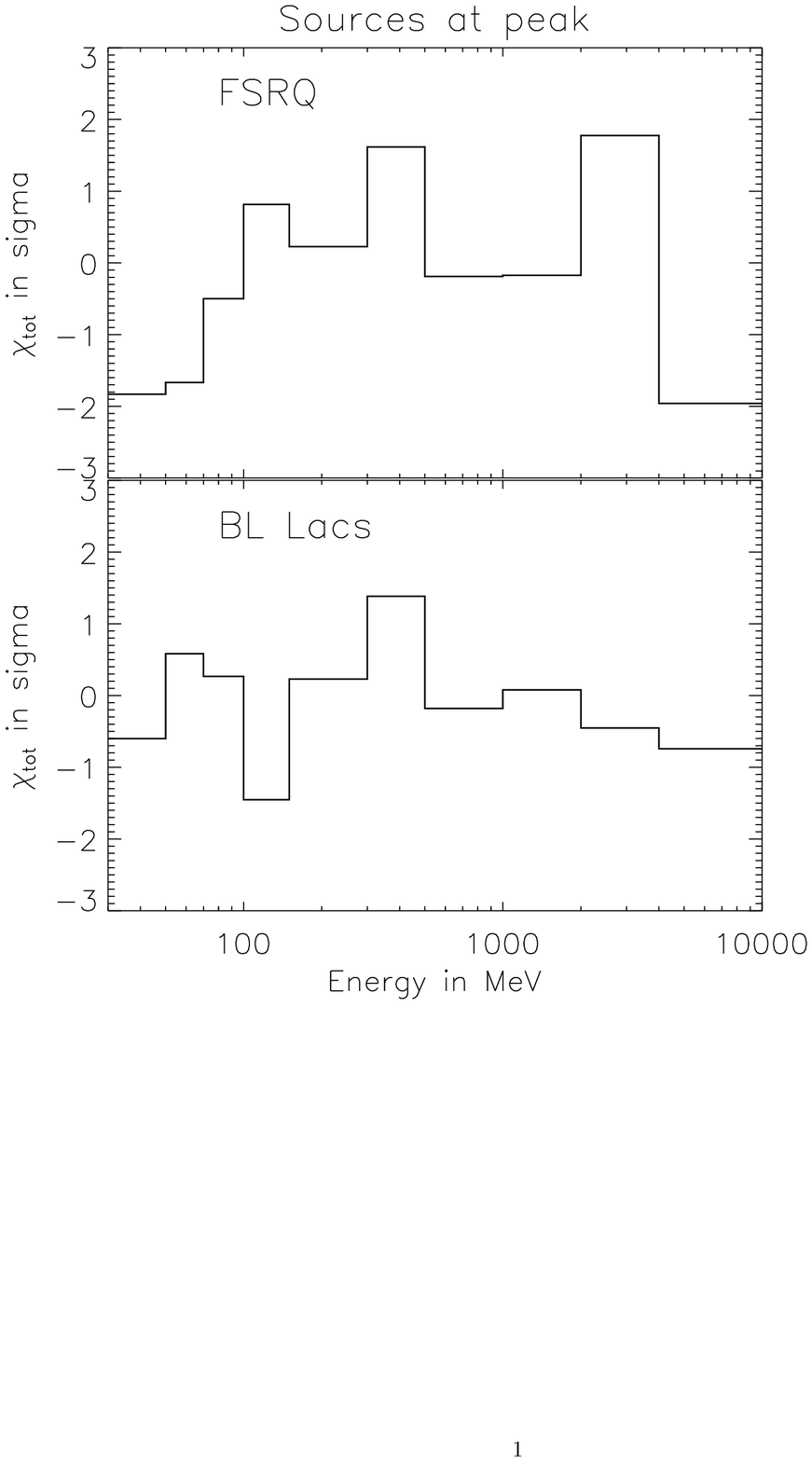,width=8.5cm,clip=}
\figure{5}{The summed power-law fit statistic of the 28 variable
FSRQs and 6 variable BL Lacs at the time of the highest flux level.
There are deviations from power-law behaviour in the peak
FSRQ spectrum at high energies and at lowest energies. The BL Lacs
at peak do not show a signature of spectral structure.}
\vskip0.2truecm}

We also see that at energies below 70 MeV and
at energies above 4 GeV the peak spectra show some evidence of a cut-off
which is more prominent in summed fit statistic than in the summed intensity.
It should be noted that the error bars for the summed intensity are mainly
determined by sources with poor statistics, i.e. short exposures or large
off-axis angles. These sources usually contribute very little to the summed
fit statistic since both at low and high energies the uncertainty
in counts exceeds
the number of detected counts. Thus the $\chi_i$ that goes into the summation in Eq.1 is practically restricted to $-1 \ll \chi_i \le 1$.
This is different for sources with good 
statistics for which even at highest energies the number of
observed and expected counts is roughly between 0.5 and 10. Thus the summed
fit statistic is dominated by sources with good statistics and its results
can not be compared directly to the summed intensity spectra.
 
At high energies a cut-off would be the expected behaviour
since the strength of opacity
effects depends directly on the flux level.
Though the low-energy deficit is significant at 50-70 MeV the average
flux {\it per source} at flare state is still a factor of 2.0$\pm$0.6 higher
than in the time-averaged case (cf. Fig.1). The flux per source at
30-50 MeV is the same as in the time-averaged case. Thus it appears that
the \gr flux below $\sim$ 50 MeV does not change, or at least not
in phase with the \gr flux above 100 MeV.

If the \gr flaring spectrum of FSRQ is indeed a power-law above 70 MeV,
which is then cut off at a few GeV due to increasing opacity, we may
underestimate the significance level of this cut-off in our analysis.
The reason is that the power-law fit tries to fit both the power-law part
and the cut-offs, and therefore will underestimate the power-law
part at medium energies which is in principle the null hypothesis in the
statistical test for cut-offs. This effect can be observed in Fig.5 as 
a strong preponderance of positive deviations between 100 MeV and 4 GeV.

To get a better idea of the significance level of the cut-offs we have
repeated the power-law fits for the peak phases of FSRQs 
under the constraint that now the fit is based on the energy band of 70 MeV
to 4 GeV and then extended to calculate the true deviations in the outer
energy bands. The result is shown in Fig.6. At energies below 70 MeV there
is a deficiency of intensity compared to power-law behaviour with total
statistical significance of 3.0$\sigma$ while at high energies above 4 GeV
we observe an intensity deficit with 2.6$\sigma$ significance. As mentioned
before the statistical significance has to be inferred from the summed fit 
statistic and it cannot be easily compared to the summed intensity spectrum.

\vskip0.4truecm
{\psfig{figure=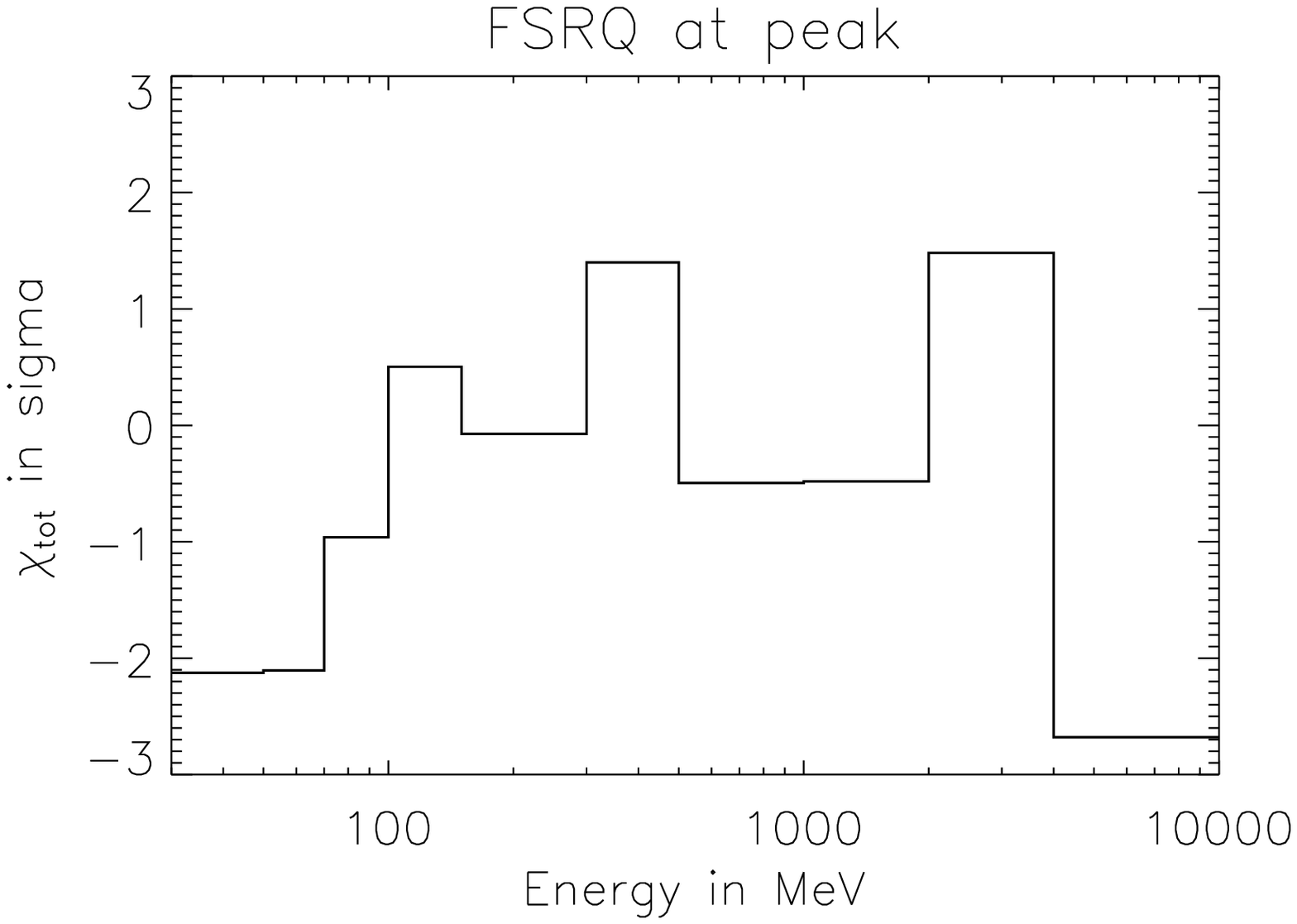,width=8.5cm,clip=}
\figure{6}{The summed power-law fit statistic of the 28 variable FSRQ
at the time of their peak flux level. In contrast to Fig.5 here the
power-law fit is based only on the data between 70 MeV and 4 GeV, i.e.
omitting the outer energy bands for which a deviation was suspected.
The total statistical significance of the spectral breaks is 3.0$\sigma$ at
low energies and 2.6$\sigma$ at high energies.}
\vskip0.2truecm}

This result is stable against the choice of source. The highest flux level of a
source does not necessarily imply the highest significance. We have tested
whether our analysis is influenced by data with poor statistics by excluding
all sources with formal significance of less than 6$\,\sigma$ in the
viewing period of peak flux. There was no change in the summed fit statistic
beyond small statistical fluctuations. On the other hand, some sources
have strong secondary maxima in their \gr light curves or have been observed 
at similar flux levels in adjacent viewing periods. To do the opposite test
we have extended our data base by including all observations which yielded 
an integrated flux of $S\ge 10^{-6}\ {\rm cm^{-2}\,sec^{-1}}$ or for
which the derived flux was within 2$\,\sigma$ of the peak value. Again our
conclusion remained unchanged. The problem with the latter test is that though
the statistical basis improves, individual sources, which appear three or
four times, start to influence the result. The fact that our result is indeed
robust and does not depend on a specific selection criteria, gives
us confidence that it is real.

\titleb{Possible Systematics}

Pulsar data show that at low energies the effective area of EGRET needs
to be decreased from the value determined by the pre-flight calibrations.
The accuracy of this correction factor can be estimated to
be better than 30\% at 30 MeV to 50 MeV and better than 10\% at 50 MeV to
70 MeV (Fierro 1995, $\S$3.4).
There is another source of uncertainty at low energies
which has to do with spill-over. At \gr energies below 100 MeV EGRET's
effective area decreases rapidly (Thompson et al. 1993). Hence photons
with original energy of around 100 MeV, which are misidentified as
of lower energy, get a strong weight and mimic a high intensity at
low energies. For the typical $E^{-2}$ spectrum around 10\% of the low
energy photons are due to this spill-over. Standard EGRET software
for spectral fits accounts for this spill-over as far as the relation
between counts and intensity is concerned. However this effect also
influences the effective point-spread function at low energies which is
calculated under the assumption of an $E^{-2.1}$ spectrum. Since the possible
error in the point-spread function is around 10\% we expect corresponding
uncertainties in the likelihood fits of similar order. In total we estimate
less than 50\% systematical uncertainties in the intensity level at lowest \gr
energies and around 20\% in the second energy band
which is to be compared to the total statistical uncertainty in the
summed intensity spectrum of quasars in Fig.4 where we have 130\% at 30 MeV
to 50 MeV and 32\% at 50 MeV to 70 MeV. In both cases the statistical uncertainties are much larger than the systematic uncertainties so that the
former are a fair measure of the total uncertainty at low \gr energies.

At high \gr energies the spill-over does not play a strong r\^ ole, since
EGRET's effective area changes only weakly with energy in this range.
Nevertheless possible calibration errors of effective area and
point-spread function
may influence the result. Also if a significant fraction of the \gr sources
is misidentified, then by using the positions of the radio sources we may
underestimate the number of observed photons at high energies where the
point-spread function is most narrow. Here it is instructive to compare
Fig.6 to the corresponding result for the average emission of FSRQs
in Fig.3. There we had no signal for a high-energy cut-off. 
We have tested this also for the case that the power-law fit is derived
only between 70 MeV and 4 GeV (as in Fig.6) with similar result: there is not
even a 1$\sigma$ indication for a cut-off beyond 4 GeV. Since any systematic
effect should influence also the average FSRQ spectrum we can be
sure that the observed high energy cut-off in the peak FSRQ
spectrum is real. A misidentification of the \gr sources with 
FSRQ in general, such that the \gr spectra discussed here are not those
of FSRQ, has been found extremely unlikely in statistical studies (Mattox et
al. 1997).

We have tested the reliability of our method by Monte-Carlo simulations.
These simulations would detect systematic problems in the analysis tools,
which may arise from the small photon numbers both at low and
at high \gr energies. We did not detect significant systematic deviations
from a Gaussian distribution of the variable $\chi_{tot}$ of Eq.1.

\titlea{Discussion}
\titleb{The deficit at low energies}

We have seen in the previous section that in case of FSRQ
the \gr spectra deviate from power-law below 70 MeV. This is most significant
with 3.0$\sigma$ when the emission of variable sources at flare
state is considered. In the time-averaged spectrum the effect is less
pronounced which indicates that if there is a deficit in the average
spectra, it will be weaker since we have better statistics there.

The extrapolation of X-ray data or the OSSE spectra for many sources imply
a spectral break at a few MeV energies. These breaks have to be extended
features regardless of their origin, even if an annihilation feature is
superimposed. This is because the standard radiation processes
like inverse Compton scattering are not monochromatic, when the target photon
distribution is neither beamed or monochromatic. So even if there were a sharp
break in the electron spectrum, the corresponding break in the \gr spectrum
would be smeared out. It is, however, questionable whether
this is sufficient to account for the observed deficit below 70 MeV, which is
a factor 10 higher in energy than the typical break energy. EGRET's 
statistics are limited in this energy range, and hence it needs some effort
to produce such a strong signal at 50-70 MeV. We would then also expect
a signal at 70-100 MeV where the statistics are much better.

We therefore prefer to interprete our result in the sense that the
\gr emission of FSRQ between a few hundred keV and a 
few GeV is not a one-component spectrum, but rather the superposition
of different emission processes. 

If the jet material consists of pair plasma we may expect annihilation
and lepton-lepton bremsstrahlung to dominate the spectrum up to a few
tens of MeV while inverse-Compton scattering is more efficient at higher
energies (B\"ottcher and Schlickeiser 1995). The comptonisation part
of the spectrum is related to the electron spectrum at high particle
energies where the physical reaction timescales are shortest. This may
explain why at flare state we see a deficit below 70 MeV. 

In case the jet is made of ordinary matter we have to consider bremsstrahlung
and Coulomb interactions besides the inverse Compton scattering.

Generally the observed behaviour can also be caused by a low-energy
cut-off in the electron injection spectrum (B\"ottcher and Schlickeiser 1996).

\titleb{The cut-off at high energies}

BL Lacs are defined as having only weak, if any, optical lines.
That translates to them having only little backscattering material in the 
vicinity of the central machine. Hence we can expect different environmental 
conditions in BL Lacs and quasars, as far as opacity is concerned. The
main source of opacity -- photon-photon pair production -- requires
a sufficient number of target photons of the energy
$$\epsilon \ge {{2 \left(m_e\, c^2\right)^2}\over {(1-\mu)\, E_\gamma}}\ 
\eqno(2)$$
where $\mu$ is the collision angle cosine in the observers frame,
and $E_\gamma$ is the energy of the \gr to be absorpted.

Standard calculations show that the interaction of \grs with themselves
is reduced due to the equivalent of the Klein-Nishina effect, which applies
when the 
target photon energy is very much larger than the limit given by Eq.2
(Pohl et al. 1995, Dermer and Gehrels 1995). Interactions of \grs with
the X-ray continuum may affect the total \gr spectrum. What we observe,
a cut-off only at highest energies, looks more like an additional
target photon source coming into play. This could be an accretion disk,
accretion disk emission backscattered by thermal clouds, or synchrotron
photons produced in the jet itself.

For \grs with dimensionless energy $\epsilon_1$
the optical depth for pair production by collisions with ambient photons
of energy $\epsilon$ is
$$\tau_{\gamma\gamma}=2\pi \int dz \int d\mu\ (1-\mu) \int_{{2}\over {\epsilon_1 (1-\mu)}}
d\epsilon\ \sigma\, n_{ph}\ \eqno(3)$$
where
$$\sigma= {{3\, \sigma_T}\over {16}} (1-\beta^2) \left[(3-\beta^4)
\ln{{1+\beta}\over {1-\beta}} -2\beta (2-\beta^2)\right]\ \eqno(4)$$
with
$$\beta=\sqrt{1-{{2}\over {\epsilon \epsilon_1 (1-\mu)}}}\ .$$

For the original accretion disk photon the collision angle is unfavorable
($\mu \approx 1$) so that the UV photons can only interact with \grs of 100
GeV energy or more (B\"ottcher and Dermer 1995). In the direct comptonization
of the accretion disk photons the Klein-Nishina
cut-off may lead to spectral turn-overs at around 10 GeV (B\"ottcher et al.
in prep.): that is a basic characteristic of the emission process
and can not account for cut-offs which occur only at flare state.

In case of the backscattered
accretion disk photons
$$n_{ph} = {{\eta L\, z_c}\over {6 (4\pi)^2 \epsilon\,z^3 c}} \, 
\delta(\epsilon-\epsilon_0)\ \eqno(5)$$
a target photon energy $\epsilon \approx 50\,$eV is sufficient to produce
a cut-off at a few GeV observed \gr energy.
Here the delta-function is a simple approximation of the accretion disk
spectrum with luminosity L and 
$\eta z_c/z^2 \ll 1$ is the assumed re-scattering rate with scattering
optical depth $\tau_{rs} =\eta z_c/z$.
Then
$$\tau_{\gamma\gamma} \simeq {{L_{46} \tau_{rs,-5}}\over {z_{-2}}}
\int_0^{\sqrt{1-{{1}\over {\epsilon_0 \epsilon_1}}}} d\beta\ 
{{\beta}\over {(1-\beta^2)^2}}$$ $$
\hphantom{\tau_{\gamma\gamma} \simeq} \times \  
 \left[(3-\beta^4)
\ln{{1+\beta}\over {1-\beta}} -2\beta (2-\beta^2)\right]\ \eqno(6)$$
with $z$ in units of 0.01 pc and $\tau_{rs,-5}$ in units of $10^{-5}$.
If $\tau_{rs,-5}$ is larger than unity, $\tau_{\gamma\gamma}$ increases
sharply beyond unity and produces a cut-off in the \gr spectrum at 
around 6 GeV (before redshift). However, correlations between
optical depth and \gr flux above 100 MeV can only occur when the
\gr outburst is due to an increased level of target photons, for instance an
accretion disk flare. Any variation in the spectrum of radiating electrons in
the jet will have no impact on the optical depth.

If electrons with Lorentz factors of $\sim 10^6$ do exist in the jets
(as is required for the TeV emission of BL Lacs), then these will produce 
a synchrotron continuum up to X-ray energies, which provides a large number
of target photons for pair production. The optical depth for this process
is
$$\tau_{\gamma\gamma} \simeq 4\,{{L_{47}}\over {r_{1d}\, \delta_{10}^4}}
\left({{E_{\gamma}}\over {\rm 5\, GeV}}\right)\ \eqno(7)$$
where $\delta$ is the Doppler factor of the jet in units of ten, $r_{1d}$
is the effective jet radius in light days ($3\cdot 10^{15}\,$cm) in the
jet frame, and $L_{47}$ is the $\nu \, F(\nu)$ synchrotron luminosity at
energy $2\delta^2 (m_e\,c^2)^2/
E_{\gamma}$ in units of $10^{47}\,$ ergs/sec. So if a \gr outburst is
caused by the injection of a large number of high-energy electrons,
the \gr emission above a few GeV may be self-damped. 
However, detailed simulations of cooling pair plasma including first order
and second order Fokker-Planck coefficients for all interaction processes
(B\"ottcher, Pohl, and Schlickeiser, in prep.) show that the high
end of the synchrotron spectrum is likely to be swamped by 
synchrotron-self-Compton emission, which may have a hard spectrum in the
relevant energy range so that $L_{47}$ in Eq.7 increases with $\epsilon$,
i.e. decreases with $E_{\gamma}$. In that case the cut-off would be very 
smooth and the energy at which it occurs would vary strongly for different
X-ray luminosities and Doppler factor. Thus there would be no good argument
why the bulk of FSRQ considered here should show the cut-off at roughly
the same energy.

To summarize: though opacity can in principle account for cut-offs at a
few GeV photon energy, and seems to be required to explain the correlation
with the \gr flux level, there is no clear answer as to the source of
the target photons for the photon-photon pair production.
However, our findings will further constrain models and simulations for
the evolution of pair jets. 

\titleb{The relation to the diffuse \gr background}

We have seen that the time and source averaged \gr spectrum of AGN
is softer than that of the extragalactic \gr background. The spectrum
of the average AGN is dominated by that of FSRQ, partly 
since we have three times more objects of this class than BL Lacs, partly
since individual FSRQ are on average brighter. It is interesting to
see that the average BL Lac has a spectrum which is compatible with that
of the \gr background and harder than that of FSRQ.
If unresolved AGN would indeed be responsible for the bulk of the \gr background, the
BL Lacs may have to play a stronger r\^ole than previously thought.
Here we outline a scenario which would account for this.

The BL Lacs have on average
a much smaller redshift with values between 0.031 and 0.94, while
more than 50 \% of the objects in the FSRQ class have redshifts
in excess of 1.0. This indicates that in case of FSRQ we observe a
fair range of the luminosity function directly, in contrast to the
BL Lac case where we see only the tip of the iceberg. In other words,
we expect the \gr $\log N/\log S$ distribution of BL Lacs to peak at
lower \gr fluxes than that of FSRQ. As a result the
contribution of BL Lacs to the diffuse extragalactic \gr background 
may be strong despite the small number of directly observed objects, and
hence it may be that BL Lacs provide the bulk of the \gr background.
This would of course require that FSRQ and BL Lacs do not have 
drastically different evolution properties. The reader should note that
at least for radio selected BL Lacs co-adding of undetected sources
has resulted in a excess of around 3$\sigma$ significance (Lin et al. 1997),
which provides further evidence that this class of sources does emit \grs
at lower flux levels.

There is yet another effect by which BL Lacs can contribute to the 
extragalatic \gr background. Distant BL Lacs may emit \grs at energies
higher than 100 GeV like the close-by objects seen by Whipple. These
\grs will pair produce on the low-energy extragalactic background before 
reaching us. Since the pairs will immediately comptonize the microwave 
background, the energy of the TeV \grs will go into an electromagnetic
cascade and finally reappear in the form of \grs at only a few GeV
energy (Coppi and Aharonian 1997). Depending on the luminosity
distribution between GeV and TeV emission of the average BL Lac, this 
process may dominate over the direct contribution to the GeV background
for particular redshifts $z$. This effect would results in a hardening
of the background spectrum in the EGRET energy range.

We should however not completely exclude the possibility that the
spectrum of the extragalactic \gr background is ill-determined.
The separation of background and galactic emission due to
cosmic ray interactions with thermal gas is relatively easy to do 
by correlation between intensity and the gas column density. In contrast the
expected spectrum of the inverse Compton
emission is similar to that of the extragalactic background and
the intensity varies little with position at high latitudes, so that we
we cannot exclude the possibility that part of the background
intensity at higher \gr energies is misidentified inverse Compton emission.

\acknow{We thank Markus B\"ottcher, Reinhard Schlickeiser for intensive
discussions. The present work uses standard viewing period analysis
products of the EGRET group at MPE, Garching.  
The EGRET Team gratefully acknowledges support from the following:
Bundesministerium f\"{u}r Bildung, Wissenschaft, Forschung und Technologie
(BMBF), Grant 50 QV 9095 (MPE); NASA Cooperative Agreement NCC 5-93 (HSC); 
NASA Cooperative Agreement NCC 5-95 (SU); and NASA Contract NAS5-96051
(NGC).}

\begref{References}

\ref B\"ottcher M., Dermer C.D.: 1995, A\&A 302, 37

\ref B\"ottcher M., Schlickeiser R.: 1995, A\&A 302, 17

\ref B\"ottcher M., Schlickeiser R.: 1996, A\&A 306, 86

\ref Brecher K., Burbidge G.R.: 1972, ApJ 174, 253

\ref Coppi P.S., Aharonian F.A.: 1997, ApJ, submitted, astro-ph/9610176

\ref Dermer C.D., Gehrels N.: 1995, ApJ 447, 103

\ref Dermer C.D., Schlickeiser R.: 1993, ApJ 416, 484

\ref Fierro J.M.: 1995, PhD Thesis, University of Stanford

\ref Kniffen D.A. et al.: 1996, A\&AS 120, C615

\ref Lin Y.C. et al.: 1997, ApJ 476, L11

\ref Mattox J.R. et al.: 1996, ApJ 461, 396

\ref Mattox J.R. et al.: 1997, ApJ, in press

\ref McNaron-Brown K., Johnson W.N., Jung G.V.: 1995, ApJ 451, 575

\ref M\"ucke A., Pohl M., Kanbach G. et al.: 1996, in {\it Extragalactic radio
sources}, IAU Symposium 175, Kluwer, Dordrecht, 285

\ref Mukherjee R., Bertsch D.L., Bloom S.D. et al.: 1997, ApJ, submitted

\ref Pohl M., et al.: 1995, A\&A 303, 383

\ref Punch et al.: 1992, Nat 358, 477

\ref Sch\"onfelder V., Bennett K., Bloemen H. et al.: 1996, A\&AS 120, C13

\ref Sikora M., Begelman M.C., Rees M.J.: 1994, ApJ 421, 153

\ref Sreekumar P., Bertsch D.L., Dingus B.L. et al.: 1997, ApJ, submitted

\ref Thompson D.J. et al.: 1993, ApJS 86, 629

\ref Thompson D.J. et al.: 1995, ApJS 101, 259

\ref Thompson D.J. et al.: 1996, ApJS 107 , 227

\ref Vestrand W.T., Stacy J.G., Sreekumar P.: 1996, ApJ 454, L96

\ref von Montigny C., Bertsch D.L., Chiang J. et al.: 1995, ApJ 440, 525

\endref

\end
\bye